\documentclass{revtex4}
\usepackage{graphicx}
\usepackage{fancyhdr}
\usepackage{amsmath}
\usepackage{color}
\begin{document}

\title{Probing LHC Higgs Signals from Extended Electroweak Gauge Group}

%

\author{Tomohiro Abe$^{a}$\footnote{Speaker. This talk is based on the
work done in~\cite{Abe:2012fb}. }, \ Ning Chen$^{b}$, \
Hong-Jian He$^{b, c, d}$}
\affiliation{$^a$KEK Theory Center,
Tsukuba, 305-0801, Japan}
\affiliation{$^b$Institute of Modern Physics and Center for High Energy
Physics, Tsinghua University, Beijing 100084, China}
\affiliation{$^c$Center for High Energy Physics, Peking University, Beijing
100871, China} 
\affiliation{$^d$Kavli Institute for Theoretical Physics China, CAS,
Beijing 100190, China}

\begin{abstract}
We study the effects of the extended electroweak gauge sector on the
 signal strengths of the Higgs boson at the LHC.
 Extension of the Higgs sector associate with the extension of the
 electroweak gauge symmetry. In our setup, there are 
 two neutral Higgs states ($h$, $H$) and three new gauge bosons
 ($W'^{\pm}$, $Z'$). 
 We assume the lightest scalar, $h$, is what LHC found and its mass is
 125~GeV. 
  We find the enhancement of $\mu(gg \to h \to \gamma \gamma)$.
 On the other hand, other decay processes are same as or smaller than the SM
 expectation.
\end{abstract}

\maketitle

\thispagestyle{fancy}


\section{Introduction}
On July 2012, both ATLAS and CMS groups reported that they found a new
particle whose mass is around 125~GeV~\cite{Aad:2012tfa,Chatrchyan:2012ufa}.
This particle is expected to be the Higgs boson predicted in the
standard model (SM).
Although the data are consistent with this expectation,
the signal strength of the diphoton decay mode has received attention.
ATLAS experiment has detected larger signal strength than the SM expectation.
Although the current deviations from the SM are still less than 2 sigma,
 it is intriguing to explore possible  implications for new physics
 which can explain this excess. Since this process is induced via
 one-loop diagram,  it is an ideal place where new physics can readily
 set in. 
Hence, the diphoton channel can provide an effective probe of possible
 heavy new states which hide in the loop and have not yet manifested in
 the direct productions. 

In the SM, the $W$ boson loop diagrams give the dominant contributions
to this process. Therefore, if we have one more $W$ boson, namely
relatively light $W'$, then the diphoton signal might be enhanced. 
This idea is easily modelized by extending the electroweak gauge
symmetry. The minimal extension is SU(2)$\times$SU(2)$\times$U(1).
This gauge structure contains $W'^{\pm}$ and $Z'$ as well as the SM
gauge bosons.  
It is also required to extend the Higgs sector for correct
symmetry breaking pattern. In a simple realization of it, two CP-even
scalars are predicted. One of these scalars is identified as the
observed particle at the LHC.

In this talk, we focus on the signal strength of the lighter CP-even
scalar. Other interesting phenomena in this model (such as the
perturbative unitarity structure, phenomenology of heavier CP-even
scalar) are discussed in~\cite{Abe:2012fb}.

\section{Model}
The gauge symmetry in this model is 
SU(3)$_c\times$SU(2)$_0 \times$SU(2)$_1 \times$U(1)$_2$, where SU(3)$_c$
is QCD part and others are electroweak sector. We introduce two Higgs
fields, $H_1$ and $H_2$, for the electroweak symmetry breaking. The
symmetry breaking patterns are 
SU(2)$_0 \times$SU(2)$_1 \to$SU(2)$_V$ by $H_1$,
and
SU(2)$_1 \times$U(1)$_2 \to$U(1)$_V$ by $H_2$,
where the suffix $V$ stand for the diagonal part. After both symmetry
breaking, the remnant symmetry is U(1)$_{\text{QED}}$, and the six of the
gauge fields become massive. These massive gauge bosons are $W^{\pm}, Z,
W^{\prime \pm}$, and $Z'$.
Each Higgs fields contain four real scalars, and six of them are eaten
by the gauge bosons. Remaining two scalars, we call them as
$h_1$ and $h_2$, are physical degrees of freedom. These two scalars are,
however, not mass eigenstates which we define
\begin{align}
 h = & \cos \alpha h_1 - \sin \alpha h_2, \\
 H = & \sin \alpha h_1 + \cos \alpha h_2.
\end{align}
We assume $h$ is 125~GeV and $H$ is heavier than $h$.
As we will see in later, the phenomenology of $h$ is highly depending on
this mixing angle, $\alpha$.

We introduce vector-like fermions as well as chiral fermions.
The charge assignments are summarized in Table~\ref{tab:fermion}.
%
%
\begin{table}[]
\begin{center}
\caption{Assignments for fermions under the gauge group of the present model.
In the fourth and fifth columns, the U(1)$_2$ charges and SU(3)$_c$
 representations are shown for the quarks (without parentheses) and
 leptons (in parentheses), respectively.}
\label{tab:fermion}
\vspace*{3mm}
\begin{tabular}{c||cccc}\hline\hline
&&&& \\[-3.5mm]
Fermions & SU(2)$_0$ & SU(2)$_1$ & U(1)$_2$ & SU(3)$_c$\\
\hline
&&&& \\[-3.5mm]
$\Psi_{0L}$ & $\bf 2$  & $\bf 1$ & $\frac{1}{6}$
	     $\left(-\frac{1}{2}\right)$ & $\bf 3$ ($\bf 1$) \\[1mm]
\hline
$\Psi_{1L}$ & $\bf 1$ & $\bf 2$ & $\frac{1}{6}$
	     $\left(-\frac{1}{2}\right)$ & $\bf 3$ ($\bf 1$)\\[1mm]
$\Psi_{1R}$ & $\bf 1$ & $\bf 2$ & $\frac{1}{6}$
	     $\left(-\frac{1}{2}\right)$ & $\bf 3$ ($\bf 1$)\\[1mm]
\hline
$\Psi^u_{2R}$ & $\bf 1$ & $\bf 1$ & $\frac{2}{3} (0)$ & $\bf 3$ ($\bf 1$)\\[1mm]
$\Psi^d_{2R}$ & $\bf 1$ & $\bf 1$ & $-\frac{1}{3} (-1)$ & $\bf 3$ ($\bf 1$)\\[-3.5mm]
&&&&\\
\hline\hline
\end{tabular}
\end{center}
\end{table}
%
%
After the symmetry breaking, the vector-like and chiral fermions are
mixed. 
Then the mass eigenstates given by the mixture of the chiral and
vector-like fermions. 
Due to this mixing, we can suppress potentially dangerous
contributions to $S$ parameter at tree level without making $W'$ much
heavy. In addition, the extra degrees of freedom in the fermion sector
help to enhance $\sigma(gg \to h)$. We will see this cross section
enhancement is crucial to explain the diphoton excess in this model.

At tree level, the $S$ parameter is approximately given by
\begin{align}
 \alpha_{em} S
\simeq&
 -4 \sin^2 \theta_W
\frac{m_W}{m_{W'}}
\frac{g_{W'ff}}{g_{Wff}},
\end{align}
where $g_{Wff}$ and $g_{W'ff}$ are $W$ and $W'$ couplings to the SM
fermions respectively. The later coupling is given by
\begin{align}
 g_{W'ff} 
\simeq&
 -g_1
\left(
\frac{1+r^2}{r^2} \frac{m_W^2}{m_{W'}^2}
-
\sin^2 \theta_f
\right)
,
\label{eq:W'ff}
\end{align}
where $r = \langle H_2 \rangle / \langle H_1 \rangle$, and $\theta_f$ is
the mixing angle in 
the fermion sector~\footnote{$\sin \theta_f$ is determined by heavy
fermion masses and Yukawa couplings. But its detail form is not
important here.}.
%
%
We can realize $g_{W'ff} \simeq 0$ by choosing a proper
configuration $\theta_f$ for the light fermion mass-eigenstates. 
Then $S$ parameter
is suppressed even if $m_{W'}$ is lighter than
1~TeV~\cite{Cacciapaglia:2004rb, Casalbuoni:2005rs, Chivukula:2005bn, 
Foadi:2004ps, Chivukula:2005xm}. If $m_{W'} \gg
1$~TeV, then we can not
expect $W'$ affects to the diphoton excess because such a heavy $W'$ is
decoupled from the SM sector. 
So, suppressed $g_{W'ff}$ is
suitable in our purpose, and hereafter we take $g_{W'ff} =0$ to mimic
the ideal fermion delocalization \cite{Chivukula:2005bn,
Chivukula:2005xm}. Then we get 
another advantage: We can avoid direct detection
bounds~\cite{Aad:2012dm, CMS-PAS-EXO-12-010, Aad:2012vs, Aad:2012hf,
:2012vb}, because
these bounds are derived under the assumptions that $W'$ is produced via
Drell-Yan process, which never happen if
$g_{W'ff} =0$.

Since we impose $g_{W'ff}=0$, the mixing angle in the fermion sector is
determined by the parameters in the gauge sector.
It is natural to assume $m_{W'} \gg m_{W}$, then the mixing angle is
small as long as $r \sim1$~\footnote{In $r \gg 1$ and $r \ll 1$ region,
$m_{W'}$ becomes heavier than a few TeV, and $g_{W'ff}$ is not needed to
be suppressed to satisfy both electroweak precision constraints and
direct detection bounds. However, such heavy $W'$ is almost decoupled
from the SM sector and does not help to
explain the diphoton excess. So we focus only in $r \sim 1$ region in
this talk.}, as we can see from Eq.~(\ref{eq:W'ff}).
Hence,
the main components of the SM fermions are the chiral fermions. On the
other hand, the non-SM fermions are approximately the vector-like
fermions.

In this setup, we find the $h$ couplings to the SM gauge bosons are
suppressed as a consequence of the perturbative unitarity
restoration mechanism in the longitudinally polarized gauge boson
scattering processes.  
In the SM case, the amplitude of $W_L W_L \to W_L W_L$ process is
proportional to $E^2$ if the Higgs boson contribution were absent.
The Higgs boson exactly cancel this $E^2$ terms, and perturbative
unitarity is restored, not violated at higher scale. This exact
cancellation of $E^2$ terms is guaranteed by the relations among some
couplings, so called unitarity sum rules;
\begin{align}
 4 m_W^2 g_{WWWW}
=&
 3 m_Z^2 g_{WWZ}^2
+
 g_{WWh}^2
,
\end{align}
which is a consequence of gauge symmetry and renormalizablity. 
In our model case, $W'/Z'$ and $H$ contribute to the process as well.
We find the following sum rule for the $E^2$ term cancellation in
$WW \to WW$ scattering amplitude:
\begin{align}
 4 m_W^2 g_{WWWW}
=&
 3 m_Z^2 g_{WWZ}^2
+
 3 m_{Z'}^2 g_{WWZ'}^2
+
 g_{WWh}^2
+
 g_{WWH}^2
.
\label{eq:sumrule}
\end{align}
As a consequence, the $h$ coupling to the $W$ bosons should be
smaller than the SM case due to the extra contributions from $Z'$ and
$H$; otherwise an over cancellation occurs and $E^2$ terms does not vanish.
In a similar manner, we can find the suppression of the $h$ coupling to
the $Z$ bosons.
Due to these coupling suppression, Br($h\to WW/ZZ$) is smaller than
the SM case and related signal strengths are suppressed, as we will
discuss in the next section.

\section{Production cross section and signal strengths}
%
%
%
In this section, we calculate the $h$ production cross section via gluon
fusion process, and some signal strengths for $h$.

\subsection{$\sigma(gg \to h)$}
We start off by studying the $h$ production cross section via gluon
fusion process. The cross section of this process can be enhanced
because this is a loop induced process by colored
particles and our model contains extra fermions as well as the SM fermions.

\begin{figure}[h]
\centering
\includegraphics[width=0.4\hsize]{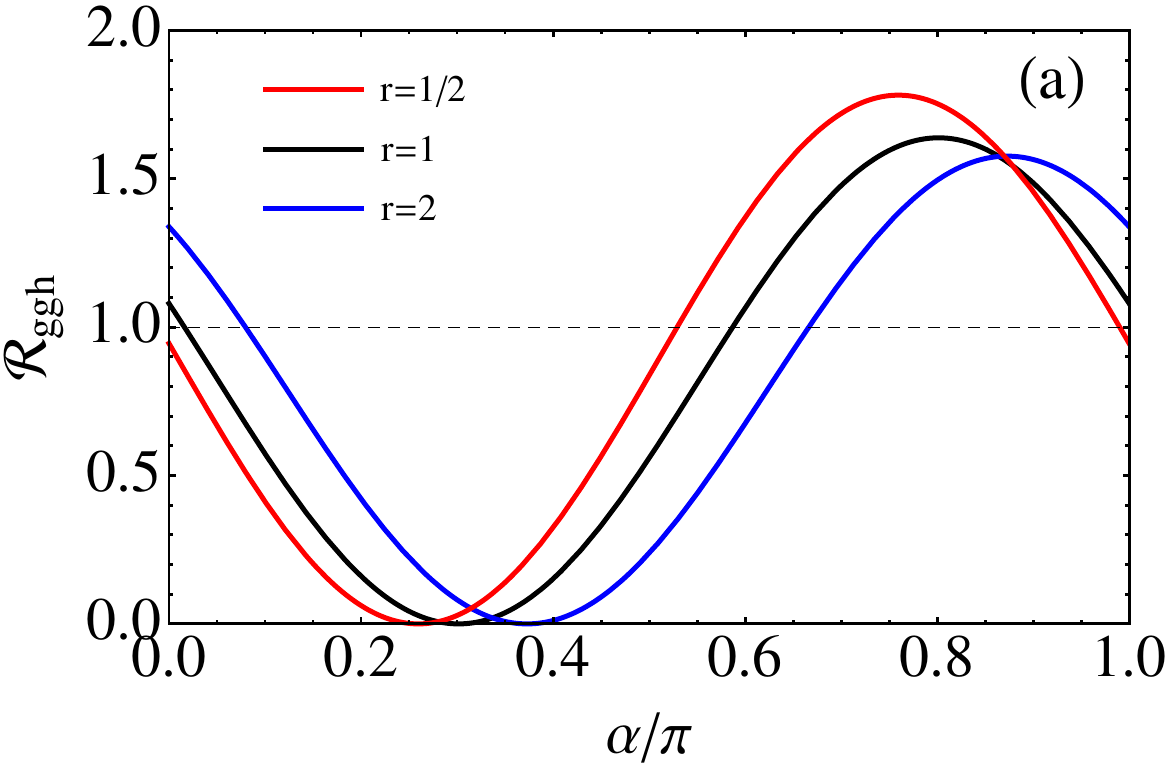}
\includegraphics[width=0.4\hsize]{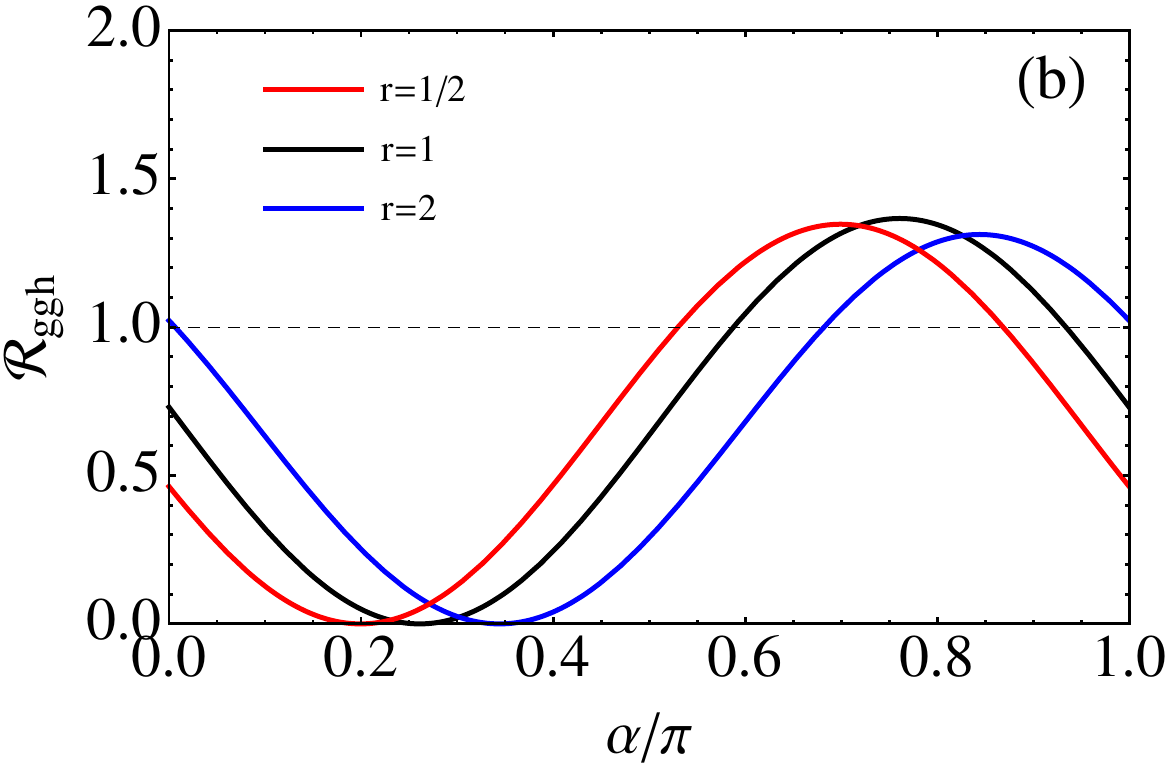}
\caption{$\sigma(gg \to h)$/SM. 
 We take $M_{W'}=400$~GeV (in the panel (a))
and $M_{W'}=600$~GeV (in the panel (b)).
The variable in the horizontal axis is the mixing angle in the Higgs
 sector, and $r$ is the VEV ratio of the two Higgs fields, $\langle H_2
 \rangle / \langle H_1 \rangle$. 
 The actual  physical space  of mixing angle $\alpha$ is $\alpha = [0,
 \pi)$, only the half of the $[0, 2\pi)$ interval. 
}
\label{fig:gg2h}
\end{figure}

In Fig.~\ref{fig:gg2h}, we show the ratio of $\sigma(gg \to h)$ in this
model to the SM one. It highly depends on the parameters in the Higgs
sector, especially $\alpha$. We find the cross section is enhanced in
some region. This enhancement is due to the contributions from the extra
fermions.
This cross section also depends on $m_{W'}$ although the
process seems independent from the gauge sector. This is due to the
condition we take, $g_{W'ff}=0$: This condition makes a connection among parameters in the
fermion sector and the gauge sector, as we can see from
Eq.~(\ref{eq:W'ff}). Hence Fig.~\ref{fig:gg2h} shows the $m_{W'}$ dependence.

\subsection{$\mu(gg \to h \to \gamma \gamma, \ WW, \ ZZ)$}
\begin{figure}[h]
\centering
\includegraphics[width=0.4\hsize]{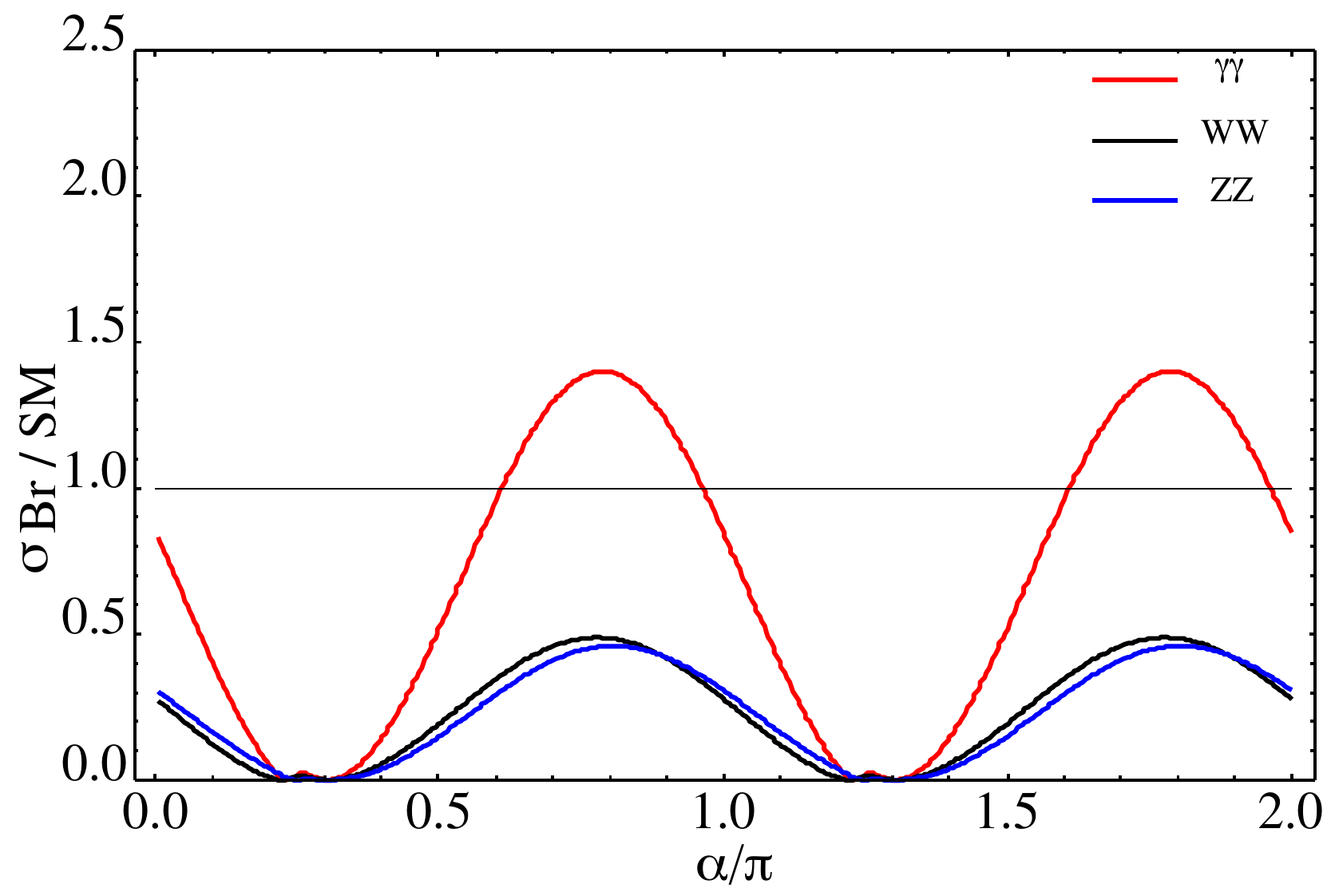}
\includegraphics[width=0.4\hsize]{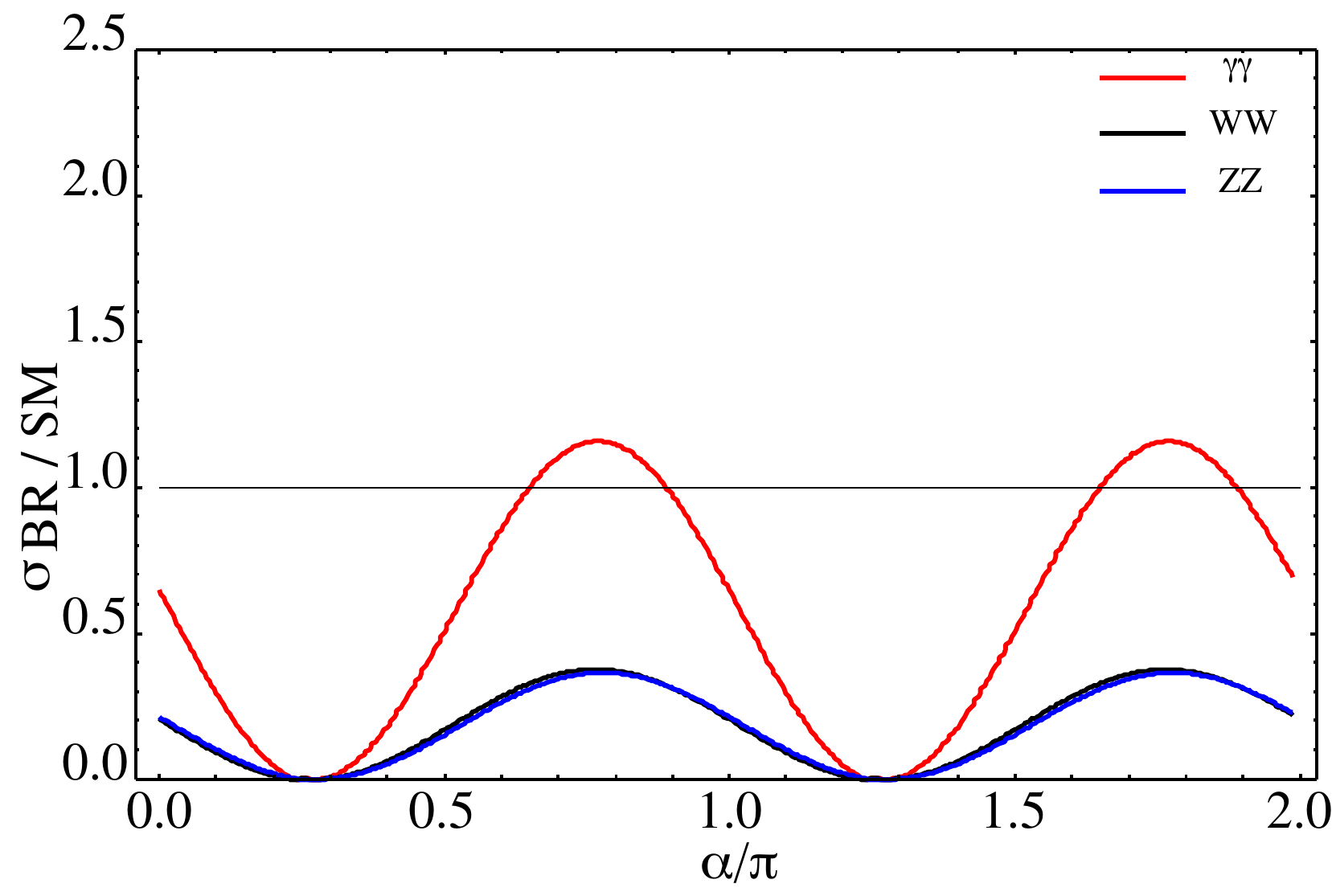}
\caption{$\mu$. 
 We take $M_{W'}=400 \ (600)$~GeV in the left (right) panel,
 and $r=1$ in both panels.
 The actual  physical space  of mixing angle $\alpha$ is $\alpha = [0,
 \pi)$, only the half of the $[0, 2\pi)$ interval. 
}
\label{fig:gg2h2VV_v1}
\end{figure}

\begin{figure}[h]
\centering
\includegraphics[width=0.4\hsize]{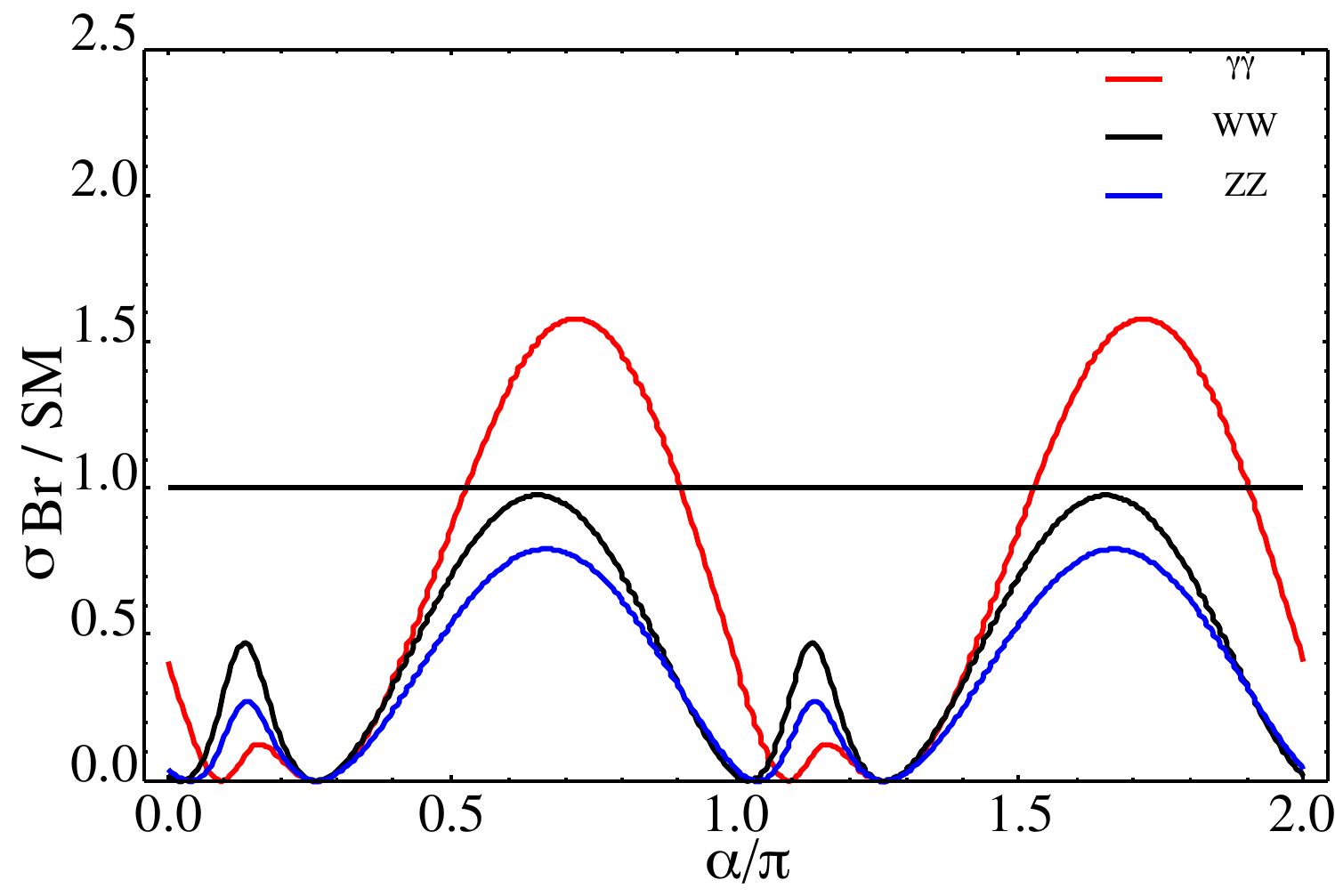}
\includegraphics[width=0.4\hsize]{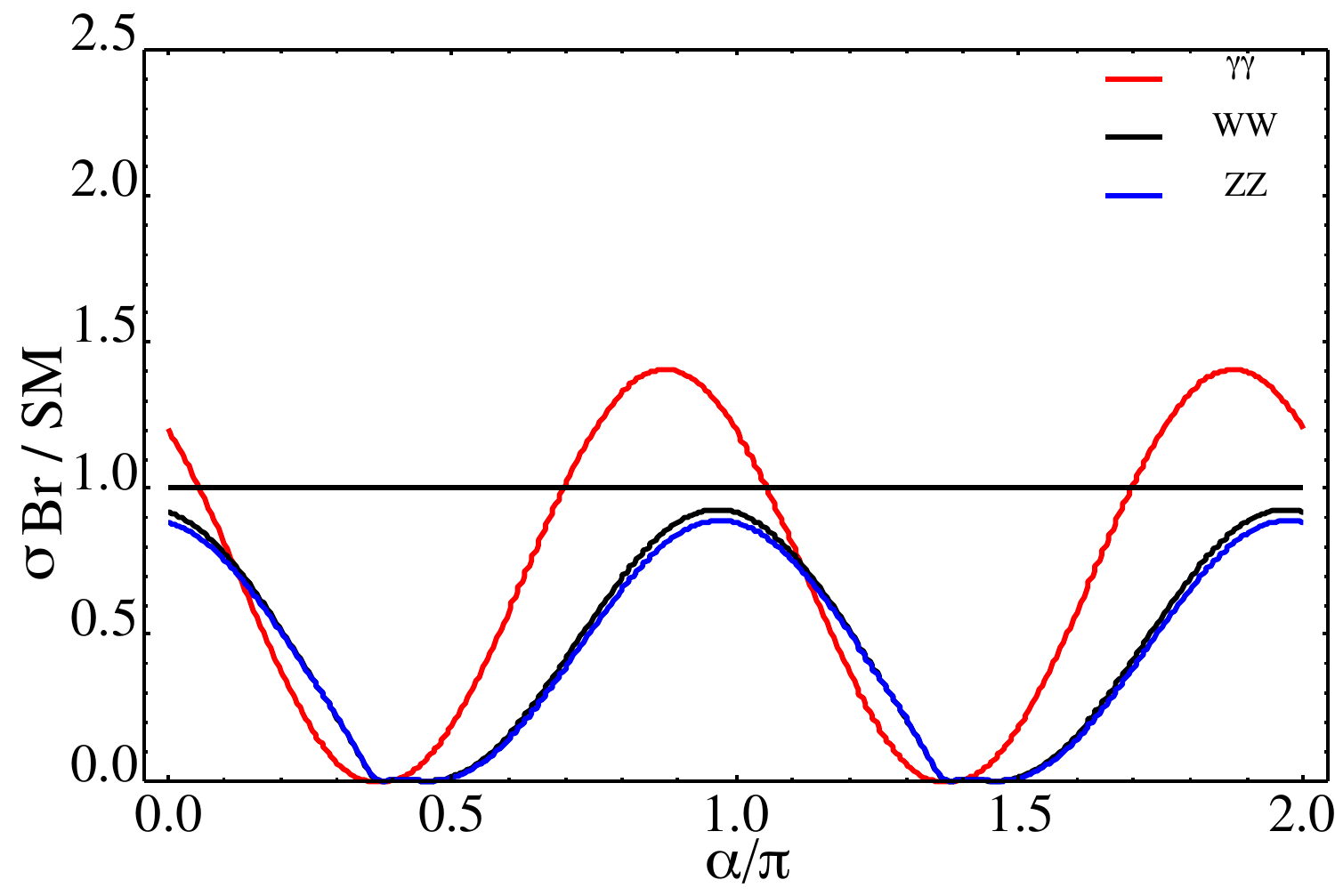}
\caption{$\mu$. 
 We take $r=2 \ (1/2)$ in the left (right) panel,
 and $M_{W'}=400$~GeV in both panels.
 The actual  physical space  of mixing angle $\alpha$ is $\alpha = [0,
 \pi)$, only the half of the $[0, 2\pi)$ interval. 
}
\label{fig:gg2h2VV_v2}
\end{figure}

Next, we calculate signal strengths, $\mu = \sigma \cdot $Br/($\sigma
\cdot$Br)$^{\text {SM}}$.  
Here we focus on three signal strengths, $\mu (gg \to h \to \gamma \gamma, \
WW, \ ZZ)$.  
In Fig.~\ref{fig:gg2h2VV_v1}, we show these signal strengths
as a function of $\alpha$, with $r=1$. We find diphoton signal excess around
$\alpha \sim 0.8 \pi$. Therefore this model can explain the excess observed by
both ATLAS and CMS.
On the other hand, $\mu (gg \to h \to WW, \ ZZ)$ is suppressed. They are
smaller than a half of the SM prediction. Since the central values of
these process are near the SM prediction, this result looks
unattractive. 
These suppression in the $WW$ and $ZZ$ channels are originated from the
suppression of the $h$ couplings to the SM gauge bosons, discussed around 
Eq.~(\ref{eq:sumrule}). Since the couplings depend on $r$,
the situation can be moderated by the choice of $r$, and we find it is true.
In Fig.~\ref{fig:gg2h2VV_v2}, we show the same plot as in
Fig.~\ref{fig:gg2h2VV_v1} but different choice of $r$.
We take $r=2 \ (1/2)$ in the left (right) panel. In these plots, we
fix $m_{W'}=400$~GeV. We find $\mu (gg \to h \to WW, \ ZZ)$ can almost
reach the SM prediction around $\alpha = 0.7 \pi ~(1.0 \pi)$ in the
left (right) panel. In those region, we still see the diphoton
excess. This is compatible with the LHC data.

\subsection{$\mu(qq' \to V h \to V ff')$ and $\mu(q_1 q_2 \to h q_3 q_4
  \to ff' q_3 q_4)$}  
\begin{figure}[h]
\centering
\includegraphics[width=0.4\hsize]{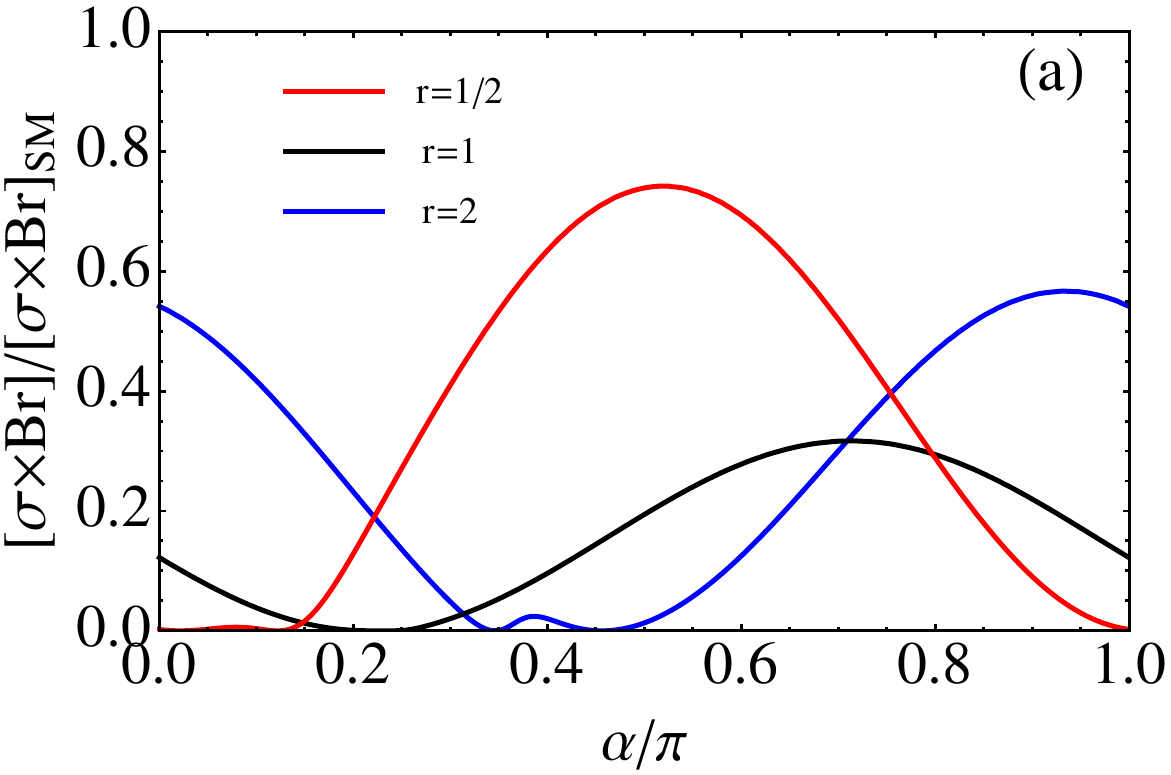}
\includegraphics[width=0.4\hsize]{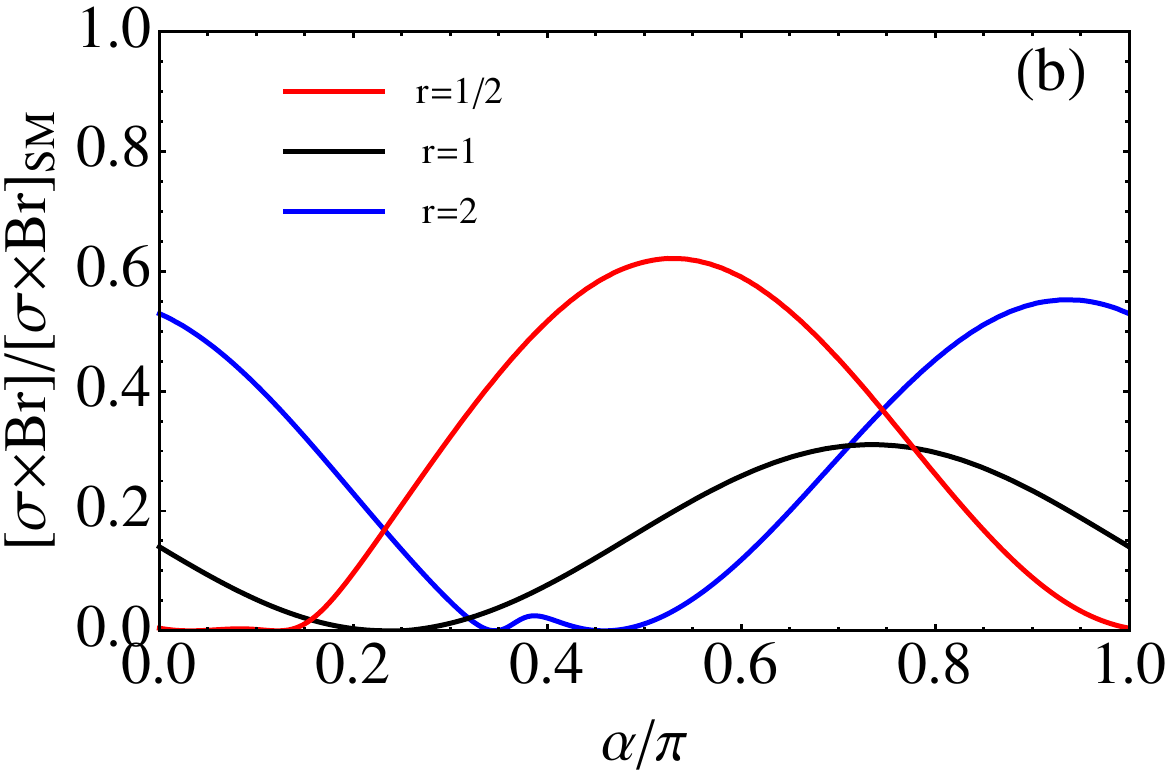}
\caption{$\mu(qq' \to V h \to V ff')$.
$\mu(q_1 q_2 \to h q_3 q_4 \to ff' q_3 q_4)$ is also same. 
 We take $M_{W'}=400$~GeV (in the panel (a)),
 $M_{W'}=600$~GeV (in the panel (b)).
 The actual  physical space  of mixing angle $\alpha$ is $\alpha = [0,
 \pi)$, only the half of the $[0, 2\pi)$ interval. 
 }
\label{fig:others}
\end{figure}

We study other important processes,
$qq' \to V^{*} \to V h \to V ff'$ and 
$q_1 q_2 \to h q_3 q_4  \to ff' q_3 q_4$,
where $V$ stands for $W$ and $Z$.
These processes are used for detecting
$h \to b\bar{b}, \tau \bar{\tau}$ process.
Since $g_{W'ff} =0$ in our analysis, the only difference of these processes
from the SM case is the $h$ couplings to the SM gauge bosons, namely
\begin{align}
\mu(qq' \to V h \to V ff')
=
\mu(q_1 q_2 \to h q_3 q_4 \to ff' q_3 q_4)
=&
\left(
\frac{g_{VVh}}{g_{VVh}^{\text {SM}}}
\right)^2
<
1.
\end{align}
The inequality in this equation is due to
the suppression of the $h$ couplings to the SM gauge bosons
as we discussed around Eq.~(\ref{eq:sumrule}), and hence this signal
strength is always suppressed.
This is a feature of this process in this model.
We explicitly show this feature by plotting the signal strength
in Fig.~\ref{fig:others}.
These two plots make it apparent that this process is certainly
suppressed compared to the SM prediction. This result is still
consistent with the current LHC data due to large statistical error, but
will become important to discriminate this model to other models in the
future.

\section{Summary}
We study the model
which has minimally extended electroweak gauge sector and vector-like
fermions. The $W'$ coupling to the SM fermions can be suppressed thanks
to the mixing between chiral and vector-like fermions. As a consequence,
$W'$ can be lighter than 1~TeV without any conflict with the $S$
parameter constraint and direct search bounds. 

In this set up, we study the diphoton signal strength of the lighter
CP-even scalar. We find enhancement of $\mu(gg \to h \to \gamma \gamma)$,
depending on the parameter choice. This enhancement is mainly due to
the enhancement of $\sigma(gg \to h)$ by the extra fermions contributions. 
On the other hand, $\mu(gg \to h \to WW/ZZ)$ can be comparable with the
SM prediction and 
be compatible with the LHC data, though they tend to be smaller than the
SM predictions. This behavior is due to the suppression of
the $h$ couplings to the SM gauge bosons. 
We also discuss the signal strength for other interesting processes,
$\mu(qq' \to Vh \to Vff')$ and $\mu(q_1 q_2 \to h q_3 q_4 \to ff' q_3
q_4)$. These processes are, for example, used to observe 
tau leptons as decay products from the Higgs boson. We find these
signal strengths are always smaller than the SM prediction. This is also
due to the suppression of the $h$ couplings to the SM gauge bosons.


\bigskip 

\end{document}